\documentclass[usenatbib]{mn2e}
\usepackage{psfig}

\newcommand{\ltaraw}{$\; \buildrel < \over \sim \;$}
\newcommand{\lta}{\lower.5ex\hbox{\ltaraw}}
\newcommand{\gtaraw}{$\; \buildrel > \over \sim \;$}
\newcommand{\gta}{\lower.5ex\hbox{\gtaraw}}

\newcommand{\hubble}{Hubble Sphere}
\newcommand{\hubbles}{Hubble Spheres}

\loadboldmathitalic   \title[The \hubble]{
How does the \hubble\ limit our view of the Universe?\footnotemark[1]} 
 \author[Lewis et al.]{Geraint F.
  Lewis$^{1}$ and Pim van Oirschot$^{2}$\\
  $^{1}$Sydney Institute for Astronomy, School of Physics, A28,
  University of Sydney, NSW 2006, Australia\\
  $^{2}$Department of Astrophysics/IMAPP, Radboud University Nijmegen, 
  P.O. Box 9010, 6500 GL Nijmegen, The Netherlands
} 
\date{\today}
\begin{document}
\maketitle
\label{firstpage}
\begin{abstract}
It has recently been claimed that the \hubble\ represents a previously unknown limit
to our view of the universe, with light we detect today coming from a proper distance less
than this ``Cosmic Horizon'' at the present time. By considering the paths of light rays in several cosmologies, 
we show that this claim is not generally true. In particular, in cosmologies dominated by phantom energy (with an
equation of state of $\omega < -1$) the proper distance to the \hubble\ decreases, and 
 light rays can cross it more 
than once in both directions; such behaviour further diminishes the claim that the \hubble\ is a fundamental,
but unrecognised, horizon in the universe.
\end{abstract}

\long\def\symbolfootnote[#1]#2{\begingroup%
  \def\thefootnote{\fnsymbol{footnote}}\footnotetext[#1]{#2}\endgroup} 

\def\newblock{\hskip .11em plus .33em minus .07em}
\section{Introduction}     \label{intro}     \symbolfootnote[1]{Research
  undertaken  as part  of  the Commonwealth  Cosmology Initiative  (CCI:
  www.thecci.org),  an  international  collaboration  supported  by  the
  Australian Research Council}
The existence of several cosmological horizons neatly carves up the 
space-time of a universe, 
with the {\it particle horizon}  containing all of the events that a particular observer 
could  ever have 
causal influence, and the {\it event horizon} containing all events that 
could ever causally influence that observer \citep{1956MNRAS.116..662R}. 
The presence and extent of these cosmological horizons depends upon the
evolution of the universal expansion, and hence ultimately on the mass-energy
content of the universe \citep[e.g. see][]{1993ApJ...406..383H}.

Recently, there have been claims of the existence of another, previously 
unrecognised horizon, dubbed the ``Cosmic Horizon" and that this fundamentally limits 
our view of the Universe \citep{2007MNRAS.382.1917M,2009IJMPD..18.1889M,2009IJMPD..18.1113M,2012MNRAS.419.2579M}.
In a spatially flat universe, this ``Cosmic Horizon'' is exactly the same as the well understood \hubble, the distance at which 
the universal expansion results in objects moving at the speed of light relative to us \citep{1991ApJ...383...60H}.
For the sake of clarity, we will henceforth assume that the universe is spatially flat and
refer to the ``Cosmic Horizon'' as the \hubble\ throughout this contribution.

In a previous paper, we demonstrated as being incorrect the claims that the \hubble\ sets a limit on
what we can observe in the universe\footnote{Just as \citet{1925JMP...4..188L} realized that 
the apparent horizon in the de Sitter metric in static form arises from a bad choice of coordinates, 
we showed that the apparent ``Cosmic Horizon'' originates from the reintroduction of these static
coordinates.} \citep{2010MNRAS.404.1633V}. 
However, \citet{2011arXiv1112.4774B} have reiterated these previous claims, considering photon paths in 
an expanding universe and stating that photons that we receive now are always from a {\it proper distance}
which is less than the present size of the \hubble. 
In this paper, we consider this claim and show that that is not generically 
true. In fact, it is possible to show that a photon may cross the \hubble\ more than once
in both directions, further revealing that
its exalted status as a previously unrecognised ``Cosmic Horizon" is still incorrect.

In Section~\ref{horizons} we discuss the key aspects of the evolution of the 
\hubble, and demonstrate how its size at the present time is not necessarily 
the limit of what we can see. We present the conclusions in Section~\ref{conclusions}.
Throughout this paper, we will consider universes described by the Friedmann-Robertson-Walker
metric.

\section{The Evolution of the Hubble Sphere}\label{horizons}

\subsection{The Hubble Sphere}\label{hubblesphere}
A consequence of the Hubble law is that objects at sufficiently large proper distance must be
receding from us at velocities greater than the speed of light. The boundary between sub- and 
super-luminal recession velocities is a spherical surface around us, known as the \hubble, and, 
by setting the recession velocity to the speed of light, $c$, into the Hubble law, is today
at a proper distance of
\begin{equation}
R_h = \frac{c}{H_o}
\label{hubbleradius}
\end{equation}
where $H_o$ is the present value of the Hubble constant.

In a generally evolving universe, the Hubble constant will be a function of time, and hence $R_h$ is 
also a function of time. In a spatially flat universe with a single component of cosmic fluid with equation of 
state, $\omega$, it is straight-forward to show that the \hubble\ evolves as
\begin{equation}
\dot{R}_h = \frac{3}{2} \left( 1 + \omega \right) c
\label{evolution}
\end{equation}
where the derivative is with respect to cosmic time  \citep{2009IJMPD..18.1113M}. In such a universe, $R_h$ clearly 
evolves at a constant rate. 

\begin{figure}
\centerline{ \psfig{figure=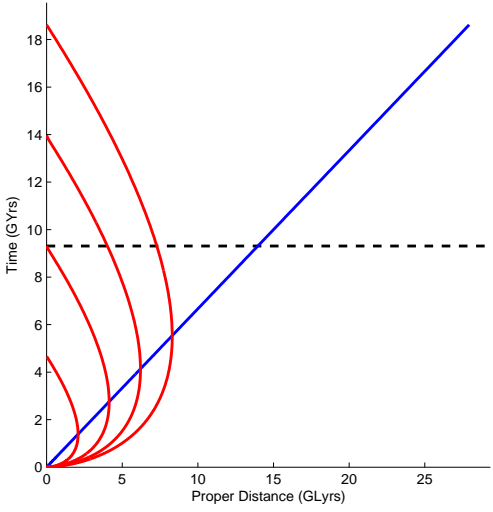,angle=0,width=2.6in}}
\caption[]{
The evolution of the \hubble\ (blue line) in an Einstein-de-Sitter universe. 
The red lines correspond to photon paths (null geodesics), while the black dashed
line is the present age of the universe (assuming $H_o=70\ {\rm km/s/Mpc}$).
\label{fig1}}
\end{figure}

\subsection{Expanding \hubbles}\label{expanding}
An example of a universe governed by Equation~\ref{evolution} is presented in Figure~\ref{fig1}, showing 
the evolution of $R_h$ (blue line) over cosmic time for an Einstein-de-Sitter universe (spatially flat, containing
only matter, so $\omega=0$). The horizontal dashed line corresponds to the present age of the universe (assuming
$H_o = 70\ {\rm km/s/Mpc}$).

The red lines in Figure~\ref{fig1} correspond to photon paths from the Big Bang (at the origin) to us
at various epochs of cosmic time; note, therefore, that these figures are essentially the same as those in 
\citet{2011arXiv1112.4774B}, but reoriented, and showing both multiple light paths and the evolution of $R_h$.
The claim by these authors is that $R_h$ today (where the blue and black dashed lines intersect, $\sim 14\ {\rm GLyrs}$) is
larger than the maximum proper distance achieved by a photon arriving at us today (roughly half that distance). 
Looking at the continually increasing value of $R_h$ into the future, and the paths of photons received into 
the future, this appears to be true.

It is important to understand, however, what Figure~\ref{fig1} is really telling us. In terms of proper distance, 
photons travel away from us at the Big Bang out to a maximum distance, before turning around and
travelling back to the origin. The point at where the photon turns around in its journey is precisely where
it crosses $R_h$; this makes intuitive sense as it can be envisaged that, due to universal expansion, this is 
the point where the photon is effectively at rest with respect to us (this is shown rigorously in Section VIII of 
\citet{1993AmJPh..61..883E}).

\begin{figure}
\centerline{ \psfig{figure=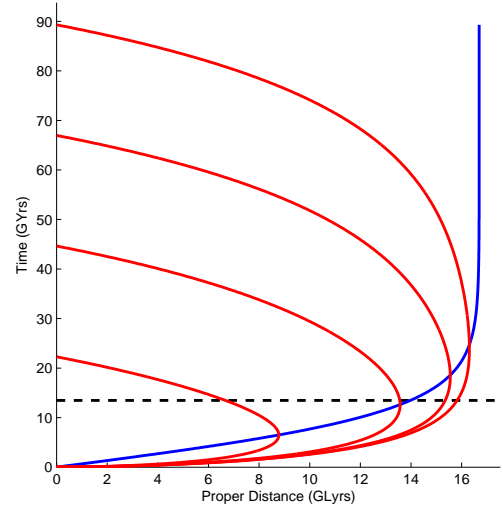,angle=0,width=2.6in}}
\caption[]{As Figure~\ref{fig1}, but for a concordance cosmological model, 
with $\Omega_m=0.3$, $\Omega_\omega=0.7$ and an equation of state of
dark energy of $\omega=-1$ (i.e. Einstein's cosmological constant).
\label{fig2}}
\end{figure}

The currently favoured cosmological model is constrained by multiple observations and contains a mix of 
cosmic fluids, being comprised of $\sim30$\% matter and $\sim70$\% dark energy
with an equation of state, $\omega\sim-1$
 \citep[e.g.][]{2003ApJS..148..175S}. The evolution of $R_h$ in such a universe is not simply described 
 by Equation~\ref{evolution}, but at early epochs, when the universe was matter-dominated (with $\omega=0$)
 we expect an evolution similar to an Einstein-de-Sitter universe, whereas at later times, the 
 universe becomes dark energy dominated. If the equation of state of dark energy is $\omega=-1$ (a cosmological
 constant),  Equation~\ref{evolution} reveals that $R_h$ is at a fixed proper distance from us.
 
 Figure~\ref{fig2} presents the evolution of $R_h$ in this universe, possessing the expected forms at early 
 and late times, with a transition period (which we are now in)\footnote{We note that such a figure is not new, and 
 an interested reader is invited to examine the excellent depiction of this cosmology in Figure~1 of 
 \citet{2004PASA...21...97D}, presenting key features of the universe in several coordinate systems.}.
 The behaviour of light rays in this cosmology is not too dissimilar to that presented in Figure~\ref{fig1}, with light
 rays traveling outwards from the Big Bang, before turning back as they cross $R_h$. A key difference, however, is
 at late times where $R_h$ asymptotes to a fixed distance from us, so that light rays will spend more and more 
 time changing direction and moving back to the observer; it is at these later times that the \hubble\
 coincides with the event horizon, truly limiting what we can see. 
 
 Again, the argument made by \citet{2011arXiv1112.4774B} appears to hold, with photons 
 arriving today not having travelled outside the present dat \hubble.
  For future observers, this remains true with all
 photons travelling out from the Big Bang to $R_h$ before heading back to the origin.

\subsection{Collapsing \hubbles}\label{collapsing}
In the examples presented in Section~\ref{expanding},  $R_h$ continually expands, to infinity in Figure~\ref{fig1} 
and asymptoting to a finite value in Figure~\ref{fig2}. But does $R_h$ have to expand? An examination of 
Equation~\ref{evolution} reveals that if the equation of state, $\omega < -1$, then $\dot{R}_h$ can be negative;
with such an equation of state, such a cosmic fluid is known as {\it phantom energy}. 
The presence of phantom energy has a dramatic effect on the expansion of the universe, potentially
resulting in a cosmic doomsday where galaxies, planets and eventually atoms are ripped apart by
the accelerating expansion \citep[e.g.][]{2003PhRvL..91g1301C}.

\begin{figure}
\centerline{ \psfig{figure=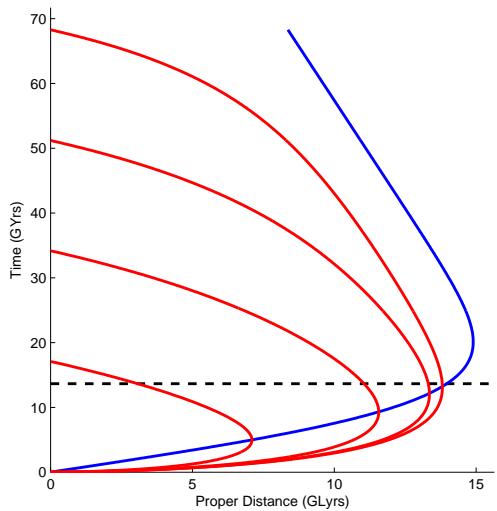,angle=0,width=2.6in}}
\caption[]{As Figure~\ref{fig2}, but adopting an equation of state of 
dark energy to be $\omega=-1.1$ (i.e. phantom energy). 
\label{fig3}}
\end{figure}

Figure~\ref{fig3} presents the evolution of $R_h$ in a universe with a present day matter density of
$\Omega_m=0.3$ and dark energy density of $\Omega_\omega=0.7$, and equation of state of 
dark energy of $\omega=-1.1$. During the earlier matter-dominated stage of the universe, the 
behaviour is similar to that seen in Figure~\ref{fig2}, but as the universe becomes dark energy 
dominated, $R_h$ reaches a maximum extent and then begins to decrease.

Examining the photon paths in Figure~\ref{fig3} reveals a similar behaviour to the previous figures, 
with photons heading out from the Big Bang before turning back towards the origin, with the turning point
being when the photons cross $R_h$. Again, photons we receive today  turn around at a distance less
than $R_h$ today, as proposed by \citet{2011arXiv1112.4774B}. However, it is clear that observers in the
distant future receive photons that turned around in their journey at a proper distance substantially 
larger than $R_h$ at the time the photon is received; this directly contradicts the ideas proposed 
by  \citet{2011arXiv1112.4774B}.

\begin{figure}
\centerline{ \psfig{figure=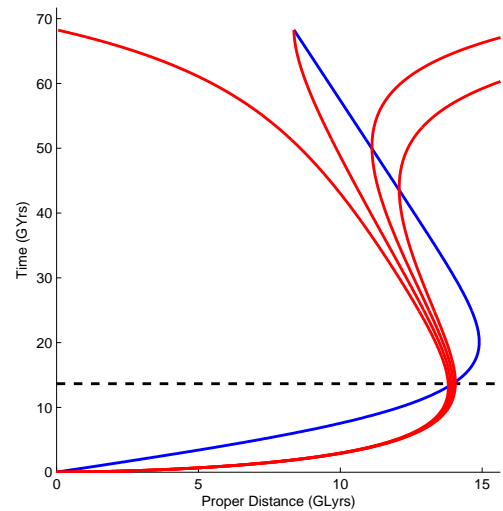,angle=0,width=2.6in}}
\caption[]{As Figure~\ref{fig3}, but considering photon paths that do not
necessarily reach the origin. 
\label{fig4}}
\end{figure}

Finally, in Figure~\ref{fig4} we further examine this phantom energy cosmology by presenting
photon paths that do not necessarily arrive back at the observer at the spatial origin. As in the previous 
figures, light paths move out from the Big Bang and turn back towards toward the observer by
passing through $R_h$. While one of the photon paths arrives at the observer, the collapsing 
\hubble\ influences the remaining photon paths, with each of them encountering $R_h$ for a
second time (and again the photon can be thought of as at rest with respect to us), before heading 
to larger proper distance. The fact that such a photon path can pass through the \hubble\ multiple times
in differing directions is another nail in the concept that the \hubble\ is a ``Cosmic Horizon".

\section{Conclusions}\label{conclusions}
In this letter, we have examined the evolution of the \hubble, $R_h$, over cosmic time, 
showing that its present size is not necessarily a limit on the maximum proper distance 
from which we are receiving photons at the present time, contrary to the claims recently
made in the literature \citep{2011arXiv1112.4774B}. 

It should be remembered that the \hubble\ is not a complex concept, and as demonstrated here 
(and in \cite{1993AmJPh..61..883E}), as well as being the boundary between sub- and super-luminal 
expansion in the universe, it represents the inflection points on a photons path between the Big
Bang and an observer (when viewed in terms of proper distance and cosmic time).  

The evolution of $R_h$ depends ultimately on the mass-energy content of the universe. 
In universes like our own, which have so far been matter dominated for much of their history, 
$R_h$ initially evolves like an Einstein-de-Sitter universe, and as $R_h$ keeps growing, it
is trivial to say that the proper distance to the turning point of a photon, 
which is equal to $R_h$ at the time of turning, is smaller than the \hubble\ now. 
For universes with a different mix of cosmic fluids, or those which components that 
evolve, such a statement cannot be necessarily made. 

We finally reiterate that photons can cross the \hubble\ multiple times, and one could imagine
a universe with an evolving dark energy component that oscillates between matter and phantom
energy. With such a universe, the \hubble\ could also oscillate in and out, with a photon 
path from the Big Bang traversing \hubble\ multiple times before reaching an observer. If
our inflationary epoch was driven by phantom energy \citep[e.g.][]{2006PhLB..632..597C},
this may have already happened. Hence the \hubble\ is not a ``Cosmic Horizon".

None of this should really come as a surprise, as the evolution of the particle and event horizons, 
and the \hubble\ have been the focus of several classic papers 
\citep[e.g.][]{1956MNRAS.116..662R,1991ApJ...383...60H,1993AmJPh..61..883E}. 
Recent contributions have added little to our understanding.

\section*{Acknowledgments} 
We would like to thank the referee, Martin Hendry, for positive comments that
improved the paper.
GFL acknowledges support from ARC Discovery Project DP0665574. 
PvO thanks the University of Sydney for hosting him during this Masters research.

\end{document}